\documentclass[prd,preprint,tightenlines,floatfix,showpacs,preprintnumbers,nofootinbib,eqsecnum]{revtex4}
 \usepackage[dvips,final]{graphicx}
  \usepackage{amssymb}
   \usepackage{amsmath}
    \usepackage{epsfig}
     \usepackage{bm}

\newcommand{\ga}{\gamma}

\newcommand{\si}{\sigma}

\begin{document}
\thispagestyle{empty}
\preprint{\hbox{}}
\vspace*{-10mm}

\title{Split Light Higgsino from Split Supersymmetry}

\author{G.~M.~Vereshkov}
\email{gveresh@ip.rsu.ru}
\author{V.~I.~Kuksa}
\author{V.~A.~Beylin}
\affiliation{%
Institute of Physics, Rostov State University,
    344090 Rostov-on-Don, Russia}%

\author{R.~S.~Pasechnik}%
\email{rpasech@theor.jinr.ru}%
\altaffiliation[\\Also at ]{Faculty of Physics, Moscow State University,
119992 Moscow, Russia}
\affiliation{%
Bogoliubov Laboratory of Theoretical Physics, JINR,
   141980 Dubna, Russia}%

\date{\today}

\begin{abstract}

New version of the MSSM scales is discussed. In this version
$\mu\ll M_{SUSY}\sim M_0\sim M_{1/2}$, where $\mu$ is the Higgsino
mass, $M_0$ is the mass scale of sleptons and squarks, $M_{1/2}$
is the mass scale of gaugino. Renormalization group motivation of
this MSSM version is proposed. Analysis of Split Supersymmetry
ideas in this case together with the Dark Matter arguments results
in the statement that the formation of residual neutralino
concentration occurs in the high symmetric phase of cosmological
plasma. The value of Higgsino mass is estimated. The recharging
process for high energy neutralinos in the neutralino-nucleus
scattering is considered. There has been reported the possibility
to check-up of the model predictions at modern experimental
facilities NUSEL and GLAST.

\end{abstract}



\maketitle

\section{Introduction}

Our optimism to find in the LHC experiments some new dynamical
effects at the $TeV$-scale is based, in particular, on the MSSM
features. As the simplest supersymmetric generalization of the SM,
the MSSM provides not only UV stability of the Higgs bosons mass
spectrum at $M_{SUSY}\sim 1 \;TeV$, but leads to unification of
the gauge couplings at $M_{GUT} \gg M_{SUSY}$ and contains an
elegant interpretation of neutralinos as the main DM components.
Besides, and this is important, the MSSM didn't contradict to
known experimental data up to energies $\sim 1 \; TeV$. Then it's
naturally to suggest that the MSSM inevitably should make an
evident of new dynamical effects and objects at the LHC
~\cite{5Mitsou}, \cite{3Abd}, \cite{4Iash}. The "fine-tuning"
absence, which is called as "naturalness" of the MSSM spectra,
simultaneously with other features of the model, has led to the
"Great Desert" existing between $M_{SUSY}$ and $M_{GUT}$. However,
in the last time there appears certain reasons to have doubts
about the singling out of the MSSM energetic spectra.

Namely, in ~\cite{A-HDim} there was formulated an assertion that
the "naturalness", as a way to avoid the "fine-tuning", can be
eliminated, in particular, in the framework of multi-vacua ideas
from superstring theories. It leads to the Split Supersymmetry
Model where $M_{SUSY} \gg 1 \; TeV$ and all superscalar masses are
much higher than $M_{EW}$ except of Higgs doublet, which is
retained at the electro-weak (EW) scale. The gauge coupling
unification as the most important feature of any reasonable SUSY
model formulation ~\cite{GiuRom} is conserved in the Split
Supersymmetry at sufficiently high scale $M_{GUT}$ to provide the
proton stability. A careful analysis of various aspects of such
type models, where Supersymmetry is split from the EW scale, is
considered in ~\cite{AHDimRev}. Some experimental consequences of
the Split Supersymmetry ideas are investigated in
~\cite{nunez},~\cite{kilian},~\cite{masiero}.

Because of any inner dynamical reasons absence, only from the
one-loop renormgroup analysis of the SUSY $SU(5)$ containing the
MSSM, there is the chance to populate "the Great Desert" with some
intermediate scales ~\cite{ourRG}. The "population" allows to
conserve the gauge couplings' unification at $M_{GUT}\ge 10^{15}
\; GeV$ ~\cite{OurHep}. The RG analysis, carried out in our work,
has shown that there are just two possible variants. The first one
is the Light Gaugino scenario that is well investigated early
\cite{aloisio1}, \cite{gondolo1}, \cite{gondolo2}, \cite{pierce}.
The second one is the Split Light Higgsino scenario, that is the
object of our investigations. The Split Light Higgsino scenario is
a SUSY variant, where Higgs bosons can be kept at $M_{EW}$ scale,
all color degrees of freedom should be much heavier than the
electro-weak ones, unification of gauge couplings occurs at $\sim
10^{15} \; GeV$ and protects the proton stability. The mass scales
of all superpartners, except Higgsino, lies in the region $\sim
10^4 \; TeV.$ It is important that in the models with the Split
Supersymmetry the Higgsino mass can remain arbitrary and can be
placed not far from the EW scale. In our paper some properties of
this model that has a Split Supersymmetry structure are
considered.

Obviously, models with Split Supersymmetry naturally lead to the
damping of scalar quarks, leptons and gauginos contributions near
EW scale. Another important consequence of the Light Higgsino
splitting is nearly degeneration in mass of neutralino and closest
chargino \cite{GiuRom}, \cite{AHDimRev}, \cite{OurHep},
\cite{Mizuta} (note, that experimental searches for the
degeneration were carried out at LEP energies ~\cite{OPAL} without
any evident signals). The masses of the lowest states (neutralinos
and charginos) and their interaction features are the key
questions for the experimental data predictions and
interpretation. All Split Supersymmetry Models are "unnatural",
and we need in some additional sources of information about the
SUSY with drifted to higher scales mass spectra.

As it was noted in ~\cite{GiuRom}, ~\cite{OurHep} the DM nature is
just one more "link between new physics and EW scale". In the
current work we have joined the Split Supersymmetry ideas with the
assumption that Dark Matter has to be composed of neutralino as
LSP. Split Light Higgsino model leads to specific predictions on
the LSP properties and, consequently, on the DM structure. The
Lightest Supersymmetric Particles (LSP) in this model have a
structure of Higgsino. We developed analysis of the residual
neutralino concentration in the cosmological plasma. It is shown
that formation of the DM concentration occurs in the high
symmetric phase of the cosmological plasma. This fact allowed us
to estimate the value of Higgsino mass $M_{\chi}\approx 3\, TeV$
within an error following from experimental uncertainties in DM
mass density determination.

Because of strong degeneration mass of chargino and neutralino in
our model the detection them at the LHC will be impossible. So
supersymmetry with the gap between Higgsino states and Higgs
bosons, on one side, and scalar states and gauginos, on the other
side, has some features that can be established in specific
experiments only. Due to particular properties of Split Light
neutralino and important role of neutralino in the DM physics, we
will see below that such investigations are possible at NUSEL
~\cite{NUSEL}, where direct observations of galactic neutralino at
underground detectors are planned, and at the satellite detector
GLAST where there are the possibilities to catch some indirect
signals from neutralino annihilation in the Galactic halo.

We have shown that the neutralino-nucleon cross section is
spin-dependent. Analysis of neutralino-chargino recharge on
nucleons is presented. This effect can take place if neutralinos
with kinetic energy $\sim 100\,GeV$ are attending in the cosmic
rays spectrum. The annihilation of Galactic neutralinos occurs in
$Z^0Z^0$ and $W^{+}W^{-}$ pairs in $t$- and $s$-channels, and also
in lepton-antilepton and quark-antiquark pairs in the $s$-channel.
The contribution to the annihilation spectrum from first two
processes is calculated with using experimental data about hadron
multiplicities at the mass surfaces of $Z^0,W^{\pm}$-bosons; the
calculation of quark-antiquark $s$-channel contribution is based
on the phenomenological model of hadron multiplicities at
$\sqrt{s}=2M_{\chi}$ with using negative binomial distribution.

The paper is organized as follows. In Section 2 the one-loop
renormalization group analysis of the SUSY $SU(5)$ containing the
MSSM is fulfilled and it is shown that there are just two possible
variants of supersymmetry: the Light Gaugino scenario and the
Light (Split) Higgsino scenario. In section 3 we have extracted
from the theory some specific information about properties of LSP
in the framework of Light (Split) Higgsino scenario. The chargino
and neutralino mass spectrum is dominated by radiation
corrections. It is quasi-degenerated and calculable exactly
without renormalization. Section 4 devoted to studying of
neutralino residual concentration and its evolution in the
cosmological plasma. It is shown here that the formation of
residual neutralino concentration occurs in the high symmetric
phase of cosmological plasma. In section 5 the possibilities for
direct and indirect searches of Dark Matter in the case of Light
(Split) Higgsino scenario are considered. Section 6 contains some
concluding remarks.

\section{Renormgroup analysis}

Standard and well known MSSM is motivated by two arguments: 1) the
used hierarchy of scales leads to convergence of all MSSM
couplings at some $M_{GUT}$ that is compatible with experimental
restriction from the proton lifetime; 2) the scale of SUSY
breaking is not very high, $M_{SUSY}\sim M_{1/2}$, so quadratic
divergencies are compensated near $M_H$, i.e. long before
$M_{GUT}$. Due to the first reason the MSSM evidently can be built
into GUT theory and into supergravity/superstring theory. But the
second argument seems not obligatory condition from the QFT point
of view ~\cite{ourRG}. Exact convergence of running couplings
takes place in the Split Higgsino Scenario too.

In \cite{A-HDim} such convergence has been shown for the SUSY
breaking scale lying in the intermediate supergauge region. In our
renormgroup analysis all $SU(5)_{SUSY}$ degrees of freedom despite
of quark-lepton states are taken into consideration involving the
states that are near $M_{GUT}$. As we'll see their contributions
are important for the choice of the SUSY scales hierarchy.

Experimentally running couplings are fixed at $M_Z$ scale:
\begin{eqnarray}
 \displaystyle \alpha^{-1}(M_Z)=127.922\pm 0.027,
  & & \quad \alpha_s(M_Z)=0.1200\pm0.0028, \nonumber \\
      \quad \sin^2\theta_W(M_Z)
  &=& 0.23113 \pm 0.00015.
\end{eqnarray}
In renormgroup equations the following values are used as initial:
\begin{eqnarray}  \vspace*{5mm}
  \displaystyle \alpha_1^{-1}(M_Z)=\frac35\alpha^{-1}(M_Z)\cos^2\theta_W(M_Z)
  &=& 59.0132 \mp(0.0384)_{\sin^2\theta_W}\pm(0.0124)_\alpha, \nonumber
\\ \vspace*{5mm}
  \displaystyle \alpha_2^{-1}(M_Z)=\alpha^{-1}(M_Z)\sin^2\theta_W(M_Z)
  &=& 29.5666\pm(0.0192)_{\sin^2\theta_W}\pm0.0062)_\alpha, \nonumber
\\
  \alpha_3^{-1}(M_Z)=\alpha_s^{-1}(M_Z)
  &=& 8.3333\pm0.1944.
  \label{exp}
\end{eqnarray}
Known equations for running couplings at one-loop level are:
\begin{eqnarray} \displaystyle
\alpha_i^{-1}(Q_2)=\alpha_i^{-1}(Q_1)+\frac{b_i}{2\pi}\ln\frac{Q_2}{Q_1},\qquad
b_i=\sum_jb_{ij}. \label{b} \end{eqnarray} In the sum all states
with masses $M_j<Q_2/2$ at $Q_2>Q_1$ are taken into account.

It is specifically that extra states should be taken into account
in these equations. Namely, singlet quarks and their superpartners
$(D_L,\,\tilde D_L),\;(D_R,\,\tilde D_R)$ are contained in
superhiggs quintets of $SU(5)_{SUSY}$, so they are in the model
inevitably (see also \cite{ourRG}). Besides chiral superfields $(
\Phi_L,\,\tilde\Phi_L)$ and $(\Psi_L,\,\tilde\Psi_L)$ in adjoint
representations of $SU(2)$ and $SU(3)$, respectively, survive from
superhiggs 24-multiplet. In the minimal $SU(5)_{SUSY}$ masses of
the states $M_5=(M_D,\,M_{\tilde D}),\;
M_{24}=(M_{\Psi},\,M_{\tilde\Psi},\,M_{\Phi},\, M_{\tilde \Phi})$
are generated by interaction with Higgs condensate at the GUT
scale, but the couplings of the interaction are phenomenological
so they are not fixed. In this scenario we evidently assume that
$M_5,\,M_{24}< M_{GUT}$ and, in principle, this inequality can be
fulfilled with accuracy in 1 -- 2 orders. Thus, for one-loop
running couplings  at the scale $q^2=(2M_{GUT})^2$ we have:
\begin{eqnarray}
 \begin{array}{c}  \vspace*{5mm}
    \displaystyle \alpha_1^{-1}(2M_{GUT})=\alpha_1^{-1}(M_Z)-\frac{103}{60\pi}\ln 2+
    \frac{1}{2\pi}\left(-7\ln M_{GUT}+\frac{4}{15}\ln
    M_{D}+\frac{2}{15}\ln M_{\tilde D}\right.
\\ \vspace*{5mm}
    \displaystyle \left. +\frac{11}{10}\ln M_{\tilde q} +\frac{9}{10}\ln
    M_{\tilde l}+\frac25\ln \mu +\frac{1}{10}\ln M_H +\frac{17}{30}\ln
    M_t+\frac{53}{15}\ln M_Z \right),
\\ \vspace*{5mm}
    \displaystyle \alpha_2^{-1}(2M_{GUT})=\alpha_2^{-1}(M_Z)-\frac{7}{4\pi}\ln 2+
    \frac{1}{2\pi}\left(-3\ln M_{GUT}+\frac43\ln
    M_{\tilde\Phi}+\frac23\ln M_{\Phi}\right.
\\ \vspace*{5mm}
    \displaystyle \left.+\frac{3}{2}\ln M_{\tilde q} +\frac{1}{2}\ln M_{\tilde
    l}+\frac43\ln M_{\tilde W}+\frac23\ln \mu +\frac{1}{6}\ln M_H
    +\frac{1}{2}\ln M_t-\frac{11}{3}\ln M_Z \right),
\\ \vspace*{5mm}
    \displaystyle \alpha_3^{-1}(2M_{GUT})=\alpha_3^{-1}(M_Z)+\frac{23}{6\pi}\ln 2+
    \frac{1}{2\pi}\left(-\ln M_{GUT}+2\ln M_{\tilde\Psi}+\ln M_\Psi \right.
\\
    \displaystyle \left.+\frac{2}{3}\ln M_{D}+\frac{1}{3}\ln M_{\tilde D}+ 2\ln
    M_{\tilde q} +2\ln M_{\tilde g}+\frac23\ln M_t-\frac{23}{3}\ln M_Z \right).
 \end{array}
 \label{gut}
\end{eqnarray}
Here $M_0=(M_{\tilde q},\,M_{\tilde l})$ -- masses of scalar
quarks and leptons averaging in chiralities and generations; $M_t$
-- $t$-quark mass; other parameters were introduced above. In
(\ref{gut}) it is supposed that $h$-boson mass lies near $M_Z$ and
other higgses $H,\,A,\,H^\pm$ are placed at the $M_H$ scale. Note
that singlet superstates and residual Higgs superfields can be
formally eliminated from (\ref{gut}), if their masses are equaled
to $M_{GUT}$ identically.

At first step all couplings have been recalculated at $2M_Z$
scale, all the SM states contribute into running of couplings
despite of $W^\pm,\,Z^0$, Higgs bosons and $t$-quark. At the time
terms with $\ln 2$ occur, which are quantitatively important for
the calculations. Between $(2M_Z,\;2M_t)$ scales the following
states emerge: $W^\pm,\,Z^0$ and one Higgs doublet containing
light $h-$boson and longitudinal degrees of freedom of
$W^\pm,\,Z^0$. At these stages $Z\bar qq$ vertex was used for
calculations of $\alpha_2^{-1}(2M_Z),\,\alpha_2^{-1}(2M_t)$. Above
the $2M_t$ scale calculations were carried out in a standard
manner. It is important that equations (\ref{gut}) do not depend
on specific arrangement of
$M_{24},\,\,M_5,\,M_0,\,M_{1/2},\,\mu,\,M_H$ degrees of freedom at
energetic scales.

Now, equaling couplings at $M_{GUT}$, from (\ref{gut}) we get
following expressions:
\begin{eqnarray}
\displaystyle
M_{GUT}=Ak_1M_Z\left(\frac{M_Z}{M'_{1/2}}\right)^{2/9},\qquad
\mu=Bk_2M_Z\left(\frac{M_Z}{M'_{1/2}}\right)^{1/3}, \label{3}
\end{eqnarray}
where
\begin{eqnarray} \vspace*{5mm}
 \displaystyle k_1
    &=& K_{\tilde q\tilde l}^{-1/12}K_{GUT1}^{1/3}\equiv
        \left(\frac{M_{\tilde l }}{M_{\tilde
        q}}\right)^{1/12}\left(\frac{M_{GUT}}{M'_{GUT}}\right)^{1/3}, \nonumber
\\
 \displaystyle k_2
    &=& K_{Ht}^{-1/4}K_{\tilde q\tilde l}^{1/4}K_{\tilde g\tilde
        W}^{7/2}K_{GUT2}^{-1}\equiv \left(\frac{M_t}{M_H}\right)^{1/4}
        \left(\frac{M_{\tilde q }}{M_{\tilde l}}\right)^{1/4}
        \left(\frac{M_{\tilde g }}{M_{\tilde W}}\right)^{5/2}
        \left(\frac{M''_{GUT}}{M_{GUT}}\right), \nonumber
 \label{4}
\end{eqnarray}

\[
 \begin{array}{c} \vspace*{4mm}
  \displaystyle M'_{1/2}\equiv (M_{\tilde W}M_{\tilde g})^{1/2},\qquad M'_{GUT} \equiv
  (M_{\tilde \Psi}M_{\tilde \Phi})^{1/3}(M_{\Psi}M_{\Phi})^{1/6}\leq M_{GUT},
\\ \vspace*{2mm}
  \displaystyle M''_{GUT} \equiv \frac{(M^2_{\tilde
  \Psi}M_{\Psi})^{7/6}(M_D^2M_{\tilde D})^{1/2}}{(M^2_{\tilde
  \Phi}M_{\Phi})^{4/3}}\leq M_{GUT},
\\ \vspace*{3mm}
  \displaystyle A=\exp\left(\frac{\pi}{18}(5\alpha_1^{-1}(M_Z)-3\alpha_2^{-1}(M_Z)-
  2\alpha_3^{-1}(M_Z))-\frac{11}{18}\ln2\right)=(1.57\times ^{1.09}_{0.92})\cdot 10^{14},
\\
  \displaystyle B=\exp\left(\frac{\pi}{3}(5\alpha_1^{-1}(M_Z)-12\alpha_2^{-1}(M_Z)+
  7\alpha_3^{-1}(M_Z))+\frac{157}{12}\ln2\right)=(2.0\times ^{0.15}_{6.56})\cdot 10^{3}.
 \end{array}
 \label{5}
\]

Here all parameters $K$ with various indexes are defined as
quantities having values more than unity. Note that
$K_{GUT1},\,K_{GUT2}$ are not under the theoretical control
neither in the MSSM nor in the $SU(5)_{SUSY}$. So we assume that
$1\le K_{GUT1}\simeq K_{GUT2}\le 10$. There is an experimental
restriction for heavy Higgs bosons \cite{Hagiwara}: $M_H>114.4
\;GeV$. It leads to following variation of $K_{Ht}$: $2\le
K_{Ht}\le 10$. Values of $K_{\tilde q\tilde l},\,K_{\tilde g\tilde
W}$ are determined from renormgroup evolution from $M_{GUT}$ to
$M_0,\,M_{1/2}$. Here we suppose that $1.5\le K_{\tilde q\tilde
l}\simeq K_{\tilde g\tilde W}\le 2.5$.

There were analyzed $M'_{1/2}$ and $\mu$ depending on $M_{GUT}$:
\begin{eqnarray}
 \displaystyle M'_{1/2}(M_{GUT})=(Ak_1)^{9/2}M_Z^{11/2}\times
 M_{GUT}^{-9/2},\qquad \mu(M_{GUT})=
 \frac{Bk_2}{(Ak_1)^{3/2}M_Z^{1/2}}\times M_{GUT}^{3/2}.
 \label{6}
\end{eqnarray}

We used known restrictions for the proton lifetime $(\tau_p\geq
10^{32}\;yr$ at $M_{GUT}\geq 10^{15}\;GeV)$ and for $M_{SUSY}$
$(M_{SUSY}\sim M'_{1/2}>100 \;GeV$ for $M_{GUT}< 3\cdot 10^{16}
\;GeV)$.

\begin{figure}[h]
\begin{minipage}{1.3\textwidth}
\epsfxsize=\textwidth\epsfbox{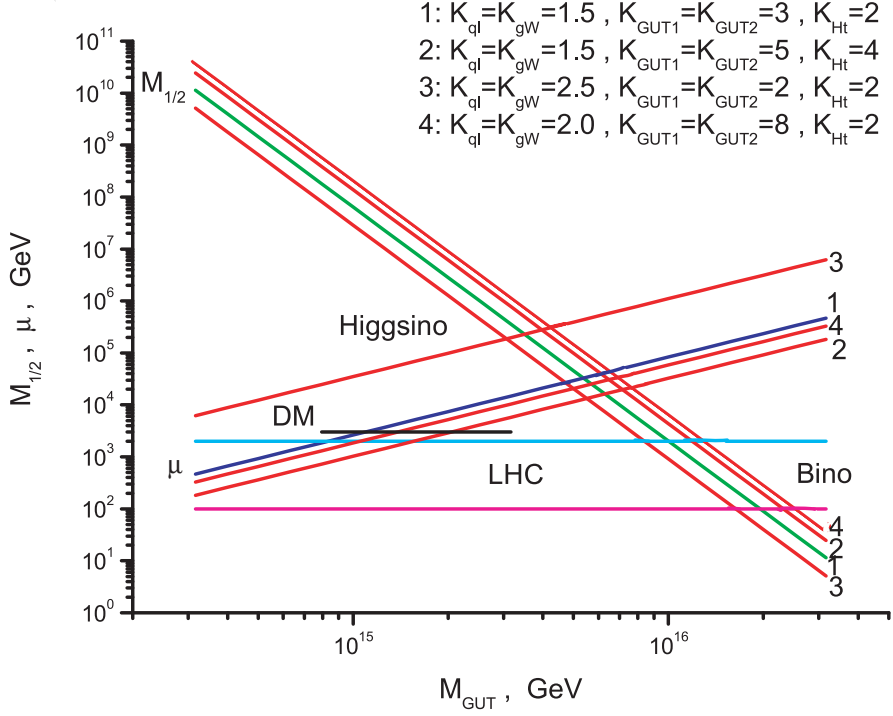}
\end{minipage}
   \caption{\label{fig:fig2} \small Two variants of the MSSM scales --
   bino-like LSP $\mu \gg M'_{1/2}$ and Higgsino-like LSP $\mu \ll M'_{1/2}$.}
\end{figure}

Two variants -- with $\mu \ll M'_{1/2}$ or $\mu \gg M'_{1/2}$ --
were found from analysis of (\ref{6}). They are shown at
Fig.~\ref{fig:fig2}.
Refusing the "Supergauge Desert" idea, nevertheless we have a
possibility to evaluate separated Higgsino mass accepting the
hypothesis that neutralino is a carrier of the DM.

\section{Objects of investigation and its properties.}

As it is known, after the EW symmetry breaking supersymmetric
partners of Higgs and electroweak gauge fields forms two Dirac
electrically charged particles -- charginos $\chi^\pm_{1,2}$, and
four Majorana electrically neutral particles -- neutralinos
$\chi^0_{\alpha},\;\alpha=1,2,3,4.$ Lightest supersymmetric
particle $\chi^0_1$ is main candidate to be the cold Dark Matter
component \cite{Primack}, \cite{Primack1}. Generally $\chi^0_1$ is
the superposition of $U(1)$ gaugino $\tilde B$ ("bino"), neutral
$SU(2)$ gaugino $\tilde{W}_{3}$ ("wino") and two Higgsinos $\tilde
h_1^0,\;\tilde h_2^0.$ Structure and properties of chargino and
neutralino depend on relations between characteristic MSSM scales.
Strong theoretical arguments in favor of one or another hierarchy
of scales are absent.

In the previous section one has shown that only two variant of
MSSM are theoretically natural. First variant -- the light gaugino
scenario (bino or wino or their mixture), in which
\begin{eqnarray}
   |\mu|\gg M_{0}\sim
   M_{1/2}\sim M_{SUSY}>M_{EW}.
 \label{gaugino}
\end{eqnarray}
-- is well investigated early \cite{aloisio1}, \cite{gondolo1},
\cite{gondolo2}, \cite{pierce}.

Now we consider the second alternative variant of MSSM, in which
the Higgsino mass scale is the nearest for EW scale, but
$M_{SUSY}$ is in the multi-TeV region -- so-called Light (Split)
Higgsino scenario (this model obviously possesses a Split
Supersymmetry property), where
\begin{eqnarray}
 \displaystyle M_{0}\sim
   M_{1/2}\sim
   M_{SUSY}\gg |\mu| > M_{EW}.
 \label{M}
\end{eqnarray}
In this model light states $\chi^0_{1,\,2}$ and $\chi^\pm_1$ have
(almost pure) Higgsino structure with masses
\begin{eqnarray}  \vspace*{5mm}
 \displaystyle M_{\chi^0_1}
  &\simeq& |\mu|-\frac{M_Z^2(1+sign(\mu)\sin
           2\beta)}{2}\left(\frac{\cos^2\theta}{M_{\tilde
           W}}+\frac{\sin^2\theta}{M_{\tilde B}}\right)\approx
           |\mu|-\frac{M_{Z}^2}{2 M_{SUSY}}, \nonumber
\\ \vspace*{5mm}
 \displaystyle |M_{\chi^0_2}|
  &\simeq& |\mu|+\frac{M_Z^2(1-sign(\mu)\sin
           2\beta)}{2}\left(\frac{\cos^2\theta}{M_{\tilde
           W}}+\frac{\sin^2\theta}{M_{\tilde B}}\right)\approx
           |\mu|+\frac{M_{Z}^2}{2 M_{SUSY}},
\\
 \displaystyle M_{\chi^\pm_1}
  &\simeq& |\mu| - \frac{M_W^2}{M_{\tilde
           W}}\left(\frac{|\mu|}{M_{\tilde W}}+ sign(\mu)\sin 2\beta
           \right)\approx |\mu|. \nonumber
 \label{H}
\end{eqnarray}
Heavier states $\chi^0_{3,\,4}$ and $\chi^\pm_2$
lie near $M_{SUSY}$:
\begin{eqnarray}
 \displaystyle M_{\chi^0_3}
 \approx M_{\tilde B},\qquad M_{\chi^0_4}
 \approx M_{\tilde W} ,\qquad
 M_{\chi^\pm_2}\approx M_{\tilde W}.
 \label{G}
\end{eqnarray}

Then all processes near EW scale are described by the SM
Lagrangian together with extra Lagrangian of light Higgsinos
interactions with photons and vector bosons:
\begin{eqnarray}  \vspace*{5mm}
 \displaystyle \Delta L
  &=& \left(eA_\mu-\frac{g_2}{2\cos\theta}(1-2\sin^2\theta)Z_\mu
      \right) \bar \chi^-_1\ga^\mu\chi^-_1 +
      \frac{g_2}{2\cos\theta}Z_\mu(\bar \chi^0_{1}\ga^\mu \chi^0_{2}
      +\bar \chi^0_{2}\ga^\mu \chi^0_{1})+ \nonumber
\\
 \displaystyle
  &+& \frac{g_2}{\sqrt{2}}W^+_\mu(\bar \chi^0_1+\bar
      \chi^0_2)\ga^\mu \chi^-_1+\frac{g_2}{\sqrt{2}}W^-_\mu \bar
      \chi^-_1 \ga^\mu (\chi^0_1+ \chi^0_2).
 \label{LH}
\end{eqnarray}
Here neutralinos are 4-component Majorana spinors.

\begin{figure}[h]
 \centerline{\includegraphics[width=0.95\textwidth]{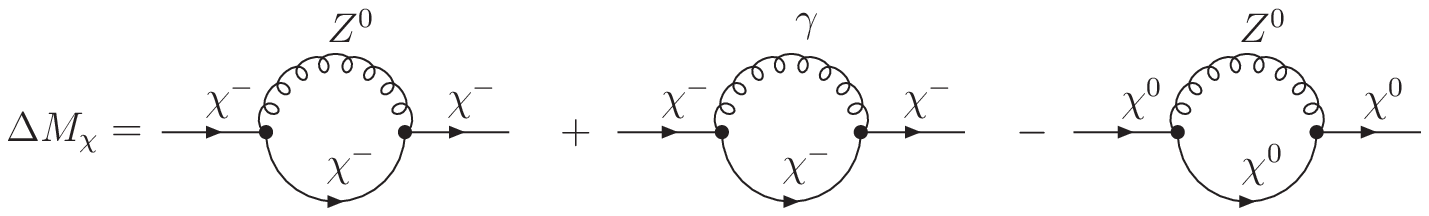}}
   \caption{\label{fig:fig1} \small One-loop diagrams contributing to the neutralino-chargino
   mass splitting.}
\end{figure}

If $M_{SUSY}>1.4\times 10^4 \; GeV$ the radiation corrections
begin to dominate in the chargino and neutralino masses formation.
Corresponding one-loop diagrams for the mass splitting are in
Fig.~\ref{fig:fig1}.
On the mass shell these mass operators take the form of the
following finite integral:
\[\Delta
M_\chi=-\frac{ie^2M_{Z}^2}{8\pi^4}\int\frac{(\hat{q}-M_{\chi})dq}
{q^2(q^2-M_{Z}^2)[(q+p)^2-M_{\chi}^2]}\;.
\]
For the case $M_{\chi}\gg M_Z$ we get:
\begin{eqnarray}
    \Delta M_\chi \simeq
    \frac{\alpha(M_{\chi}) M_Z}{2}.
 \label{delta}
\end{eqnarray}
To fix the value $\alpha$ at the scale of $M_{\chi}$ we used
$M_{H}\sim M_t\sim 200 \; GeV$ ~\cite{Hboson}, \cite{belyaev}.
Neutralino mass we estimate as $M_{\chi}\sim 3\; TeV$ (for details
see later). With a good accuracy $\Delta M_\chi \simeq 360\; MeV$.
Then for the chargino lifetime we find:
\begin{eqnarray}
    \displaystyle \tau_{\chi_1^\pm}=
    \frac{15\pi^3}{G_F^2}(\Delta M_\chi)^{-5}\simeq 0.4\times 10^{-9}\;s.
 \label{t}
\end{eqnarray}

If we refuse from Supergauge Desert we haven't theoretical
considerations on the separated Higgsino mass. Nevertheless we can
attempt to make an important step to solve this problem when we
use the hypothesis that neutralino is a carrier of Dark Matter in
the Universe.

\section{Estimation of the split Higgsino mass from cosmology}

In discussing versus of the theory physics of neutral and charged
Higgsinos fixes by the Lagrangian (\ref{LH}) and the mass spectrum
(\ref{H}) with the radiation splitting (\ref{delta}). Therefore
any quantitative predictions of the theory depend on two
parameters $\mu,\;M_H$ only. Concrete value $M_H$ isn't essential.
The basic parameter of the Split Light Higgsino model is $\mu.$ To
evaluate it let's suppose neutralino as a carrier of Dark Matter
in the Universe. Irreversible neutralino annihilation starts in
the cosmological plasma at the moment $t_0$ and the temperature
$T_0$, when mean energy of relativistic quarks and leptons
compares with neutralino mass: $\bar \varepsilon_f \simeq
3\,T_0=M_{\chi}.$ Residual density of neutralino at $t\gg t_0$
describes by the asymptotic solution of evolutional equations
\cite{gondolo}, \cite{Primack}, \cite{Primack1}:
\begin{eqnarray}
 \begin{array}{c} \vspace*{5mm}
  \displaystyle \frac{dn_{\chi}}{d t}+3Hn_{\chi}=-\frac12 n_{\chi}^2(\si v)_{ann},
\\
  \displaystyle 3H^2=8\pi GwT^4, \quad H=\frac1a \cdot \frac{da}{dt}, \quad
  wT^4=\frac{const}{a^4}.
 \end{array}
 \label{nchi}
\end{eqnarray}

Here $(\si v)_{ann}$ stands for the kinetic annihilation cross
section; $G$ is Newton's constant; $w=w(T)$ -- plasma statistical
weight for the neutralino annihilation epoch. In the SM with two
Higgs doublets the weight is $w=443\pi^2/120$ and this value is
suitable for the analysis of annihilation process at temperatures
exceeding the SM characteristic scale. For  $T < M_W$ it should be
used the value $w=23\pi^2/8$.

Time asymptotics of (\ref{nchi}) has the form:
\begin{eqnarray}
 \displaystyle
 n_{\chi}(t)=\left(\frac{a(t_0)}{a(t)}\right)^3\left(\frac{2\pi
 Gw}{3}\right)^{1/2} \frac{4T_0^2}{(\si v)_{ann}}.
 \label{nchi1}
\end{eqnarray}
Annihilation epoch begins at the moment
\begin{eqnarray}
 \displaystyle t_{0}\simeq
 \frac{1}{4T_{0}^2}\left(\frac{3}{2\pi Gw}\right)^{1/2}.
 \label{tann}
\end{eqnarray}
and ends when $T_{1}\simeq M_\chi/20$. To obtain
relic neutralino density $\rho_{\chi}=M_{\chi}n_{\chi}$ known data
on cosmological neutrino were used. At $t>t_0$ processes of quark-gluon
plasma hadronization and annihilation of $l \bar l$ pairs
practically occur through non-neutrino channels, so consequently
neutrino evolution in cosmological plasma  can be considered
adiabatically. Therefore, scale factors ratio in (\ref{nchi1}) can
be replaced by the ratio of neutrino gas temperatures:
\[
\displaystyle \frac{a(t_0)}{a(t_{U})}=\frac{T_{\nu}}{T_0}, \qquad
T_{\nu}\simeq \left(\frac{4}{11}\right)^{1/3}T_{\ga}= 1.676\times
10^{-13} \; GeV.
\]
Here $t_{U}$ stands for Universe age, $T_{\nu}\equiv
T_{\nu}(t_{U})$, $T_{\ga}$ -- the gamma relic temperature.

Finally, for the epoch when relic neutrino temperature  is
$T_\nu$, for stable neutralinos density we get:
\begin{eqnarray}
 \displaystyle
 \rho_\chi(T_\nu)=\frac{M_\chi}{T_{0}}\left(\frac{2\pi
 Gw}{3}\right)^{1/2} \frac{4T_\nu^3}{(\si v)_{ann}}.
 \label{DM}
\end{eqnarray}

From the recent WMAP data \cite{verde}, \cite{spergel} in the
modern Universe the DM mass density is:
\begin{eqnarray}
 \displaystyle
 \rho_{DM}(t_{U})\simeq (0.23\pm 0.04) \rho_c \simeq (0.94\pm
 0.34)\times 10^{-47}\;GeV^4.
 \label{DMexp}
\end{eqnarray}
where $\displaystyle
\rho_c=(4.1\pm0.8)\times 10^{-47} \;GeV^4$ is the critical density
for the Universe. Assuming the DM consists of neutralinos only, we
have equality:
\begin{eqnarray}
 \rho_{\chi}(t_{U})=\rho_{DM}(t_{U}).
 \label{ysl}
\end{eqnarray}

To estimate the value of $M_{\chi}$ from (\ref{ysl}) annihilation
kinetic cross section $(\si v)_{ann}$ vs. $M_{\chi}$ should be
obtained. But it doesn't known at which phase of cosmological
plasma the irreversible neutralino annihilation occurs, so we need
in two possible scenarios analysis.

{\bf 1. Neutralino annihilation: low-symmetric phase.}

If neutralino mass $M_\chi<3\,T_{EW}\sim 300 \;GeV$ its relic
abundance is formed after the EW transition, in low-symmetric (LS)
phase of cosmological plasma. In standard scenarios Majorana
neutralinos, which are nearly gauginos (\ref{gaugino}), annihilate
into (nearly) massless fermions and annihilation cross section
depends on the temperature essentially. In the model under
consideration neutralinos can't annihilate into fermions, so the
cross section doesn't depend on the temperature with a good
accuracy. There are two formal scenarios: annihilation of
short-lived particles ($\tau_\chi<t_0$) and coannihilation of
long-lived particles with ($\tau_\chi\gg t_0$) \cite{gondolo2}.
From our analysis it follows that variant with light split
chargino and Majorana neutralinos exists only when s-channel
annihilation is forbidden. In other words second neutralino state
$\chi_2^0$ could be unstable with the lifetime smaller than
$10^{-10}\;s$. This is possible if $M_{SUSY}\ge 10^{5} \;GeV$
only. However this scenario leads to the necessity of a neutralino
mass to the fine tuning near $M_W$. At the same time it results
to:
\[
M_{\chi_1^-}\sim M_{\chi_1^0}\sim M_W,
\]
that contradicts to the known experimental data \cite{Hagiwara}.
Formally in the LS phase there is a second solution, which
corresponds to $M_{\chi_1^0}\sim O(TeV),$ but it is incompatible
with the LS-phase neutralino annihilation. Thus, we haven't a
self-consistent results for split neutralino annihilation in the
LS phase.

{\bf 2. Neutralino annihilation: high-symmetric phase.}

As it is known for temperatures $T>T_{EW}\sim 100\;GeV,$ where
$T_{EW}$ stands for the electro-weak phase transition temperature,
cosmological plasma is in the high-symmetric phase (HS) and the
plasma doesn't contain any Higgs condensate. A special feature of
this phase is that all particles excepting Higgsinos are massless
(more exactly, their masses $m\ll T$). In HS-phase all physical
states are presented by chiral fermions and gauge fields $B,W_a
\,(a=1,2,3)$ quanta. Due to Higgs condensate absence neutralino
and chargino degrees of freedom join into the fundamental $SU(2)$
representation, i.e. Dirac field $\chi$. All states of the field
are dynamically equivalent and correspond to restored $SU(2)$
symmetry quantum numbers. Thus, instead of (\ref{LH}) it should be
used following Lagrangian:
\begin{eqnarray}
 \displaystyle \Delta L_{\chi}=
 \frac{1}{2}g_1B_\mu\bar\chi\ga^\mu\chi+\frac{1}{2}g_2W^a_\mu\bar\chi\ga^\mu\tau_a\chi,
 \label{LH1}
\end{eqnarray}
which is added to the SM Lagrangian written in
terms of gauge and chiral fields:
\begin{eqnarray}
 \begin{array}{c} \vspace*{5mm}
  \displaystyle L_{SM}= -\frac{1}{2}g_1B_\mu\bar l_L\ga^\mu l_L-g_1B_\mu\bar
  e_R\ga^\mu e_R+\frac{1}{6}g_1B_\mu\bar q_L\ga^\mu
  q_L+\frac{2}{3}g_1B_\mu\bar u_R\ga^\mu u_R-\frac{1}{3}g_1B_\mu\bar
  d_R\ga^\mu d_R
\\
  \displaystyle
  +\frac{1}{2}g_2W^a_\mu\bar l_L\ga^\mu\tau_a l_L
  +\frac{1}{2}g_2W^a_\mu\bar q_L\ga^\mu\tau_a q_L.
 \end{array}
 \label{LSM1}
\end{eqnarray}
Here corresponding flavour and family sums are suspected in all above terms.

When $|\mu|\gg T_{EW}$ irreversible neutralino annihilation is
governed by Lagrangians (\ref{LH1}) and (\ref{LSM1}) and occurs in
the HS-phase. In t- and s-channels all cross section of Higgsino
annihilation into gauge bosons and massless fermions were
calculated analogously to QCD calculation technology. The only
difference is in the fact that it is necessary to consider all
channels with initial and final states, which have an arbitrary
color in two dimensions -- it corresponds to the restored SU(2).
The full list of all considered channels is following:
\begin{eqnarray}
 \begin{array}{c} \vspace*{5mm}
 \displaystyle \chi\chi \to BB,\qquad \chi\chi \to W_aW_a ;
\\
 \displaystyle \chi\chi \to B^* \to l_L\bar l_L,\;e_R\bar e_R,\;q_L\bar
 q_L,\;u_R\bar u_R,\;d_R\bar d_R; \qquad \chi\chi \to W_a^* \to
 l_L\bar l_L,\;q_L\bar q_L,
 \end{array}
 \label{HS}
\end{eqnarray}

\begin{figure}[h]
 \centerline{\includegraphics[width=0.95\textwidth]{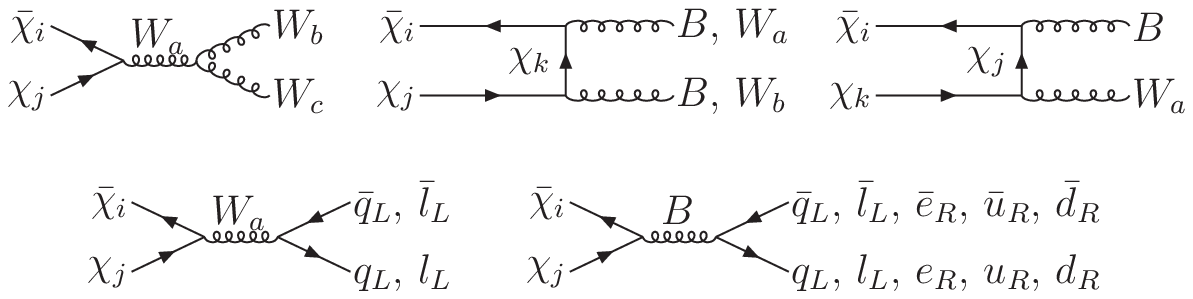}}
   \caption{\label{fig:fig3} \small Annihilation channels in the high-symmetric phase.}
\end{figure}
where $l_L,\;q_L,\;e_R,\;u_R,\;d_R$ are chiral
quarks and leptons of three generations. For total kinetic
annihilation cross section we find:
\begin{eqnarray}
 \displaystyle (\si v)_{ann}=
 \frac{21g_1^4+6g_1^2g_2^2+39g_{2}^4}{512\; \pi M_{\chi}^2}
 \label{sHS}
\end{eqnarray}
Here $g_1=g_1(2M_{\chi}),\,g_2=g_2(2M_{\chi})$ are gauge couplings
at the scale $\sqrt{s} = 2 M_{\chi}$. Assuming that the relic is
formed in the HS-phase, from theoretical expression (\ref{DM}) and
experimental data (\ref{DMexp}) we get that the DM can be made
from split neutralinos if they have the mass
\begin{eqnarray}
 M_\chi=2.9\pm 0.5 \;TeV.
 \label{Mchi}
\end{eqnarray}
Then the irreversible neutralino annihilation starts at the temperature $T_0\sim
M_\chi/3 \sim 1000\;GeV$ and finishes at $T_1\sim M_\chi/20\sim
100\;GeV$. Thus, annihilation process occurs in the HS-phase
only and it shows that the above neutralino mass estimation is
self-consistent. Evidently, the value (\ref{Mchi}) should be
treated very carefully because of the DM content is really
unknown. Moreover, for this estimation an inequality $M_\chi\ll
M_{SUSY}$ should be correct. In any case this theoretical scenario
is motivated phenomenologically no worse than known scenarios
where $M_{SUSY}\sim 1 - 2 \;TeV$.

\section{Possibilities of split neutralino observation}

Above suggested scenario with split Higgsino can't be tested at
LHC because of small chargino-neutralino mass splitting and,
consequently, too small energies of lepton pairs from these
particles' decays. In the situation the scenario check-up becomes
possible in astrophysics only.

\subsection{Direct detection}
\subsubsection{Split neutralino elastic scattering on nucleons and nuclei}

Now there are nearly twenty experimental programs for relic WIMPs
direct detection \cite{search1}, \cite{smith1}, \cite{smith2},
\cite{sumner}. Unfortunately, we haven't any clear evidence of
WIMPs existence up to now.

All operating detectors use the nuclear recoils registration
after the process of WIMPs elastic scattering at nuclei
\cite{WIMPSearch}. Separation the signal from background is a
difficult problem in these experiments and to decrease the muon
background cryogenic apparatus with the mass near 1 ton will be
used in NUSEL experiments \cite{NUSEL}. It seems that the usage
of liquid xenon as scintillator or supercooled Ge-Si crystals
are most perspective technologies. Xenon radiates when recoil
pass through it and, if Ge-Si crystals are used, it's
necessary to measure ionization energy of recoil together with
vibrational quant passed to the crystal grid. To study neutralino
observation process neutralino-nucleon elastic scattering
should be considered.

It is known that the spin-independent component of the total
cross section is due to neutralino-quark interaction by means
of scalar quarks exchange \cite{WIMPSearch}. In the considered
scenario this component is strongly damped by large scalar quark
masses. As a result there is spin-dependent component of the
total cross section only:
\begin{eqnarray}
 \si_{\chi n}=\frac{g_2^4
 m_n^2}{64\pi m_W^4} \,,\qquad \si_{\chi p}= \si_{\chi n}\cdot (1-
 4 \sin \Theta_W^2)^2.
 \label{SD}
\end{eqnarray}
For the elastic scattering off nuclei result is \cite{WIMPSearch}:
\begin{eqnarray}
 \si_{\chi nucl}\simeq
 \left(\frac{M_{nucl}}{m_p} \right)^2 \frac{4(J+1)}{3J}(\langle
 s_p\rangle+\frac{a_n}{a_p}\langle s_n\rangle)^2 \si_{\chi p}\;,
 \label{siN}
\end{eqnarray}
here $a_p (a_n)$ are WIMP-proton (neutron) couplings.

Cause we know the DM local density ($\approx 0.3 \;GeV/cm^3$) and
evaluate neutralino velocity near the Sun as $\approx 200\; km/s$,
split neutralinos flux in detector can be estimated as
$j_{\chi}\simeq 2\cdot 10^3\, cm^{-2}s^{-1}.$ In particular, one
of the most perspective elaborated detector XENON  has the
threshold $\le 10\,\;KeV$ for the energy of nuclear recoil.
Typically, the recoil energy after interaction with galactic split
neutralino is $\sim 100\;KeV$ and for non-relativistic neutralino
on xenon nuclei elastic cross section (\ref{siN}) is $\sim
10^{-35}\;cm^2.$ It gives $\sim 1 - 2$ events per hour that is
close to the planned detector background $4.5\cdot 10^{-5}\,
\;s^{-1}$ or $\sim 1$ event per hour. Thus, in the case the signal
can't unambiguously confirm the DM structure formed by split
Higgsino.

\subsubsection{Recharging of split neutralino on nucleons}

Some new processes could be possible when the relic neutralinos
penetrate through the matter. Such channels are opened:
\[\chi^0+p\to \chi^++n\,,\qquad \chi^0+n\to \chi^-+p\,,\]
and cross sections are:
\[
  \si^{*}_{\chi p}=\frac{g_2^4 |U_{ud}|^2 m_p^2}{16\pi
  m_W^4}\sqrt{\frac{E_N-\Delta m_N-\Delta M_\chi}{E_N}}\,,\qquad
  \si^{*}_{\chi n}=\frac{g_2^4 |U_{ud}|^2 m_n^2}{16\pi
  m_W^4}\sqrt{\frac{E_N+\Delta m_N-\Delta M_\chi}{E_N}}\,,
\]
here $\Delta m_N=m_n-m_p\simeq 1\; MeV$; $\Delta
M_\chi=M_{\chi_1^{\pm}}-M_{\chi^{0}} \simeq 360\; MeV$ --
chargino-neutralino mass difference that can be calculated from
(\ref{delta}); $E_N = \frac{1}{2} m_N v_{rel}^2$. An average
kinetic energy of neutralino in the locality of the Sun is about
$\sim 1\;MeV$, but recharging reaction is possible only if there
are high energy neutralinos with $E_{kin}>1\;TeV$ in cosmic rays.

\subsection{Indirect neutralino detection}
\subsubsection{Diffuse gamma ray spectrum}

\begin{figure}[h]
 \centerline{\includegraphics[width=0.95\textwidth]{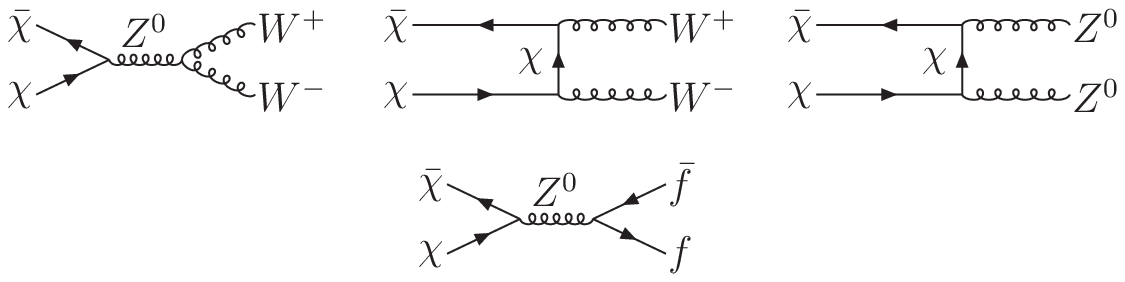}}
   \caption{\label{fig:fig4} \small Diagrams of neutralino annihilation in the Galactic halo.}
\end{figure}

Besides direct relic neutralino observation there are a set of
satellite experimental programs for the detection of neutralino
annihilation spectrum in the Galactic halo \cite{search1},
\cite{sumner}, \cite{ann}. Annihilation process and calculation of
gamma spectrum characteristic energies were evaluated in the above
scenario framework. Namely, Majorana neutralinos in the halo
annihilate into W- and Z-bosons \begin{eqnarray} \displaystyle
\chi\chi \to W^{+}W^{-},\qquad \chi\chi \to ZZ, \label{wz}
\end{eqnarray} and fermions
\begin{eqnarray} \chi\chi \to q\bar q,\; l\bar l. \label{ff} \end{eqnarray}
Corresponding diagrams are shown in Fig.~\ref{fig:fig4}.
Total kinetic cross section then is:
\begin{eqnarray}
 \displaystyle (\si v)_{ann}=
 \frac{g_2^4\,(21-40\cos^2{\theta_W}+34\cos^4{\theta_W})}{256\; \pi
 M_{\chi}^2\cos^4{\theta_W}}.
 \label{annhalo}
\end{eqnarray}

Now use the fact that total secondary hadrons multiplicity nearly
twice larger than charged hadrons multiplicity \cite{Hagiwara},
\cite{dremin}, \cite{L3}. Average multiplicity of secondary
charged hadrons $\langle \tilde n_{ch}\rangle(\sqrt{s})$ was
studied in $e^+e^-, \, pp, \, p \bar p, \, e^{\pm}p$ processes
\cite{hera}. It was established that this quantity is some
universal function of energy
\[
 \displaystyle \tilde n_{ch}(\sqrt{s})=A+B\ln \sqrt s+C\ln^2\sqrt s,\quad\tilde
 n_{ch}\equiv \langle n_{ch}\rangle (\sqrt{s}/q_0)-n_0
\]
with experimentally fixed parameters
\[
 A=3.11\pm 0.08,\quad
 B=-0.49\pm 0.09,\quad C=0.98\pm 0.02.
\]
To choice a specific channel it is necessary to fix parameters $q_0,\,n_0$. For
neutralinos annihilation $\displaystyle q_{0(\chi\chi)}=1,\;
n_{0(\chi\chi)}=0.$

Further it is supposed that the part of charged hadrons $\kappa
\equiv \langle n_{ch}\rangle /\langle n_{h} \rangle$ doesn't
depend on energy and has the value $\kappa \simeq 0.49$, which is
extracted from $Z$-peak data \cite{Hagiwara}.

For characteristic photon energies generation the following
processes are most important $\pi^0 \to 2\ga ,\;\eta^0 \to 3\ga
,\;\eta^0\to 3\pi^0 \to 6\ga$. In the annihilation channel
(\ref{ff}) total hadron multiplicity is described by following
logarithmic function with good accuracy:
\begin{eqnarray}
 \displaystyle \langle n^{ff}_h
 \rangle=\kappa^{-1}(A+B\ln 2M_\chi+C\ln^22M_\chi)\simeq
 149\,,\qquad M_\chi\simeq 3 \;TeV.
 \label{nff}
\end{eqnarray}
An average energy of neutral pions in neutralinos annihilation secondaries is
$\bar E_{\pi^0}\simeq \bar E_{\eta^0}\simeq 2M_{\chi}/\langle
n^{ff}_h \rangle$. Then in the $\pi^0\to 2\ga$ decay maximal
characteristic energy of photon is equal:
\begin{eqnarray}
 \bar E_{\ga
 (\pi^0\to\, 2\ga)}\simeq \bar E_{\pi^0}/2=20 \;GeV.
 \label{xe1}
\end{eqnarray}

Analogously, for reactions $\eta^0\to 3\ga$ and $\eta^0\to
3\pi^0\to 6\ga$ energies of photons are:
\begin{eqnarray}
 \bar E_{\ga (\eta^0\to\, 3\ga)}\simeq \bar E_{\eta^0}/3=13.5 \;GeV\,,\qquad
 \bar E_{\ga (\eta^0\to\, 6\ga)}\simeq \bar E_{\eta^0}/6=6.7 \;GeV.
 \label{xe2}
\end{eqnarray}

In the second annihilation channel (\ref{wz}) total hadron
multiplicity of $W$- and $Z$-bosons is:
\begin{eqnarray}
 \displaystyle \langle n^{WZ}_h
 \rangle\simeq 42.9.
 \label{nwz}
\end{eqnarray}
Here neutral pion average energy is $\bar E_{\pi^0}\simeq \bar E_{\eta^0}\simeq
M_{\chi}/\langle n^{WZ}_h \rangle$ and maximal characteristic
photon energies are:
\begin{eqnarray}
\begin{array}{c} \vspace*{5mm}
\bar E_{\ga (\pi^0\to\, 2\ga)}\simeq \bar E_{\pi^0}/2\simeq 35 \;GeV,
\\
\bar E_{\ga (\eta^0\to\, 3\ga)}\simeq \bar E_{\eta^0}/3=23.3
\;GeV\,,\quad \bar E_{\ga (\eta^0\to\, 6\ga)}\simeq \bar
E_{\eta^0}/6=12 \;GeV.
\end{array}
\label{xe3}
\end{eqnarray}

In a wide energy region multiplicity distribution can be
described with a good accuracy by the Negative Binomial
Distribution (NBD) that depends on energy very weakly
(logarithmically) \cite{L3}:
\begin{eqnarray}
 \begin{array}{c} \vspace*{5mm}
 \displaystyle P(n; \tilde n, k)=\frac{k(k+1)...(k+n-1)}{n!}\cdot
 \frac{(\tilde n/k)^n}{[1+(\tilde n/k)]^{n+k}}\;,
\\
 \displaystyle k^{-1}(\sqrt{s})= a+b\ln \sqrt s,
 \end{array}
 \label{nbd}
\end{eqnarray}
where $n\equiv n_{ch},\,\tilde n\equiv \tilde
n_{ch}$. For various channels coefficients in the function
$k^{-1}(\sqrt{s})$ are different:
\begin{eqnarray}
 \displaystyle a_{e^+e^-}= -0.064\pm
 0.003,\qquad b_{e^+e^-}=0.023\pm 0.002;
 \label{abee}
\end{eqnarray}
\begin{eqnarray}
 \displaystyle
 a_{pp/\bar pp}= -0.104\pm 0.004,\qquad b_{pp/\bar pp}=0.058\pm
 0.001.
 \label{abpp}
\end{eqnarray}

\begin{figure}[h]
 \centerline{\includegraphics[width=0.75\textwidth]{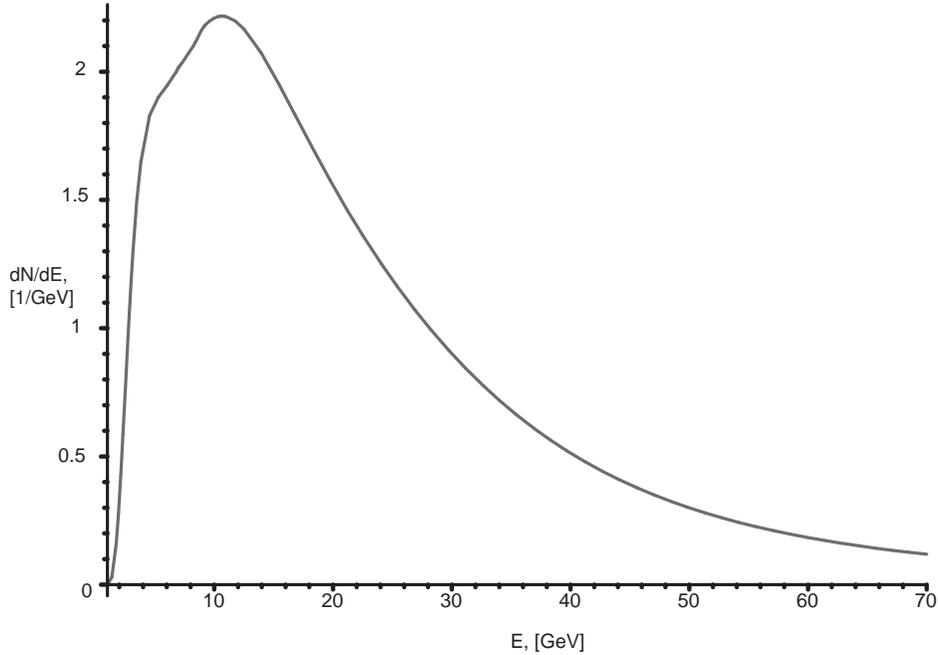}}
   \caption{\label{fig:fig5} \small Diffuse gamma ray annihilation spectrum.}
\end{figure}

Using the NBD it is possible to find approximate energetic
distribution of photons. For example, in hadronic annihilation
channel (\ref{ff}) with $Br (h)\simeq 0.58$ it is supposed that
branchings for various hadrons $Br (i/h)\equiv \langle
n_i\rangle/\langle n_h\rangle$ are nearly constants and equal to
corresponding branchings extracted from $e^+e^-$-annihilation
\cite{Hagiwara}. Each annihilating neutralino pair gives a number
of neutral pions $Br(\pi^0/h)\cdot 2\bar n P(n; \bar n, k)$ with
the probability $Br (h)$. These pions generate the following
distribution of photons $\Delta n_\ga\simeq 2\cdot
Br(h)Br(\pi^0/h)\cdot 2\bar n P(n; \tilde n, k)\Delta n$ with
energy $E_\ga=M_\chi/2n$ in the multiplicity interval $\Delta n$,
which is connected with energy interval $\Delta n=M_\chi \Delta
E_\ga/ 2E_\ga^2$. For $\eta^0$-mesons and annihilation channel
(\ref{wz}) considerations are analogous. Then the number of
photons with energy $E_{\gamma}$ per one annihilation act is:
\begin{eqnarray}
 \begin{array}{c} \vspace*{5mm}
 \displaystyle \frac{dN_\ga}{dE_\ga}\simeq
 \frac{2M_\chi}{E_\ga^2}\Large\left\{\displaystyle \left[ Br(\pi^0/h)+
 Br(\eta^0/h)\cdot Br(\eta^0\to 2\ga)\right]\times \right.
\\ \vspace*{5mm}
 \displaystyle \times \left[Br(h)\langle n^{ff}_{ch}\rangle
 P\left(\frac{M_\chi}{2E_\ga}; \langle n^{ff}_{ch}\rangle,
 k_{ff}\right)+\frac12 Br(WZ)\langle n^{WZ}_{ch}\rangle
 P\left(\frac{M_\chi}{4E_\ga}; \langle n^{WZ}_{ch}\rangle,
 k_{WZ}\right)\right]+
\\ \vspace*{5mm}
 \displaystyle+ Br(\eta^0/h)\left[Br(\eta^0\to 3\pi^0)+\frac13 Br(\eta^0\to
 \pi^+\pi^-\pi^0)\right]\times
\\ \vspace*{5mm}
 \times \displaystyle \left[Br(h)\langle n^{ff}_{ch}\rangle
 P\left(\frac{M_\chi}{6E_\ga}; \langle n^{ff}_{ch}\rangle,
 k_{ff}\right)+\frac12 Br(WZ)\langle n^{WZ}_{ch}\rangle
 P\left(\frac{M_\chi}{12E_\ga}; \langle n^{WZ}_{ch}\rangle,
 k_{WZ}\right)\right]+
\\ \vspace*{5mm}
 \displaystyle +\frac13 Br(\eta^0/h) Br(\eta^0\to \pi^+\pi^-\ga)\times
\\
 \displaystyle \left.\times \left[Br(h)\langle n^{ff}_{ch}\rangle
 P\left(\frac{M_\chi}{3E_\ga}; \langle n^{ff}_{ch}\rangle,
 k_{ff}\right)+\frac12 Br(WZ)\langle n^{WZ}_{ch}\rangle
 P\left(\frac{M_\chi}{6E_\ga}; \langle n^{WZ}_{ch}\rangle,
 k_{WZ}\right)\right]\right\}.
 \label{dN/dE}
 \end{array}
\end{eqnarray}
Here $Br(WZ)\simeq 0.2$ is the total branching for neutralino
annihilation into $W$- and $Z$-bosons; charge hadron
multiplicities $\langle n^{ff}_{ch}\rangle$ and $\langle
n^{WZ}_{ch}\rangle$ are defined in (\ref{nff}) and (\ref{nwz});
parameters $k^{-1}_{ff}=k^{-1}(2M_{\chi})=0.4$ and
$k^{-1}_{WZ}=k^{-1}(M_{\chi})=0.12$ were determined with
coefficients from (\ref{abpp}) and (\ref{abee}), respectively. The
spectrum is shown in Fig.~\ref{fig:fig5}.

Calculated value of gamma-ray flux is close to the threshold of
sensitivity for the satellite detector GLAST. Possibility of
neutralino annihilation spectrum registration at this apparatus
will be clarified when the GLAST will be put into operation.

\section{Conclusion}

The model considered (Split Higgsino Scenario) has peculiarity
that is inherent in the SM extensions having of the split mass
spectra: Higgsino mass scale is considerably lower than the soft
SUSY breaking scale: $\mu\ll M_{SUSY}\sim M_{0}\sim M_{1/2}$.

In this model, as in the standard MSSM, there is the precise
convergence of invariant couplings, the scale of convergence
$M_{GUT}$ doesn't contradict to the well known restriction on the
proton lifetime. It is essential that in the case mass spectrum of
two Majorana neutralinos and charginos is strongly degenerated
near the $\mu$ scale. For $M_{SUSY}>1.4\times 10^4 \; GeV$
radiation corrections dominate in the neutralino-chargino mass
splitting. From the mass difference calculations with the
renormgroup evolution consideration it follows
$M_{\chi_1^{\pm}}-M_{\chi_{1,2}^{0}}\simeq 360 \;MeV$ at
$M_{\chi}\simeq \mu \approx 3 \;TeV.$ The last value have got from
the analysis of the neutralino relic abundance in cosmological
plasma. Such heavy neutralinos, being of the DM carriers, should
annihilate in the high-symmetric phase forming the relic. Charged
Higgsino lifetime in the model is $\tau_{\chi_{1}^{\pm}}\simeq
0.4\times 10^{-9} \;s.$ These results and peculiarities are
inevitable because of hierarchy of the model scales and
discriminate the model considered from other models possessing of
the Split Supersymmetry property.

These physical consequences of the model  can be verified, in
principle, in some experiments planned at underground laboratory
NUSEL or satellite detector GLAST. Namely, at NUSEL it is possible
to search relic split heavy Higgsino. Due to the fact that spin
dependent part dominates in the neutralino-nucleon scattering,
neutralinos can be registered in the XENON detector. The evaluated
neutralino signal is close to the rated value of background.
Moreover, if there are high energy neutralinos $E_{\chi}\ge 1
\;TeV$ in cosmic rays the recharging process for neutralino-nuclei
scattering can be seen.

The neutralino annihilation spectrum can be observed at GLAST
detector. Diffuse radiation spectrum has been calculated and
analyzed for the galactic DM consisting of heavy (split)
neutralinos.  Contributions to the gamma ray spectrum from t- and
s-channels annihilation into $Z^0Z^0$ and $W^{+}W^{-}$ pairs were
calculated with the using of experimental data on hadron
multiplicities on $Z^0,\,W^{\pm}$ mass shell;  for quark-antiquark
s-channel contribution phenomenological model of hadron
multiplicity at $\sqrt{s}=2M_{\chi}$ was used too.

Thus, it has been shown some other possibility of the $M_{SUSY}$
scale shift to far energy region -- in comparison with the known
Split Supersymmetry variant. This splitting of the mass spectrum
is a consequence of the extra supergauge symmetries existing at
intermediate scales, between the SM and GUT. If results of the
SUSY search at LHC will be discouraged, the proposed model (or
analogous -- with other energy scales arrangement) leave the
chance to keep supersymmetry ideas, but at higher energy scale
than it is within reach of the LHC. Evaluation of $M_{SUSY}$ turn
out to be possible if the split sector of neutral and charged
Higgsino is realized. With the interpretation of neutralinos as
main component of the DM, it becomes feasible to give some
predictions for future experiments at NUSEL and GLAST.


\begin{thebibliography}{100}

\bibitem{5Mitsou}
V. A.~Mitsou, ATL-CONF-2000-002 (2000).

\bibitem{3Abd}
S.~Abdulin et al., CMS-NOTE 1998/006.

\bibitem{4Iash}
I.~Iashvili, A.~Kharchilava,  Nucl. Phys. B \textbf{526}, 153-162
(1998).

\bibitem{A-HDim}
N.~Arkani-Hamed, S.~Dimopoulos, JHEP 0506, 073 (2005).

\bibitem{GiuRom}
G. F.~Giudice, A.~Romanino, Nucl. Phys. B \textbf{699}, 65-89
(2004).

\bibitem{ourRG}
V.~Beylin, G.~Vereshkov, V.~Kuksa, in Proceedings of 16 Int.
Workshop on Quantum Field Theory and High Energy Physics, Moscow,
2001, p.300.

\bibitem{AHDimRev}
N.~Arkani-Hamed, S.~Dimopoulos, G. F.~Giudice, A.~Romanino, Nucl.
Phys. B \textbf{709}, 3-46 (2005).

\bibitem{nunez}
L.~Anchordoqui, H.~Goldberg, C.~Nunez, Phys. Rev. D\textbf{71},
065014 (2005).

\bibitem{kilian}
W.~Kilian, T.~Plehn, P.~Richardson, E.~Schmidt, hep-ph/0507137.

\bibitem{masiero}
A.~Masiero, S.~Profumo, P.~Ullio, Nucl. Phys. B\textbf{712},
86-114 (2005).

\bibitem{OurHep}
G.~Vereshkov, V.~Kuksa, V.~Beylin, R.~Pasechnik, hep-ph/0410043.

\bibitem{Drees2}
M.~Drees, M.~Nojiri, D.~Roy, Y.~Yamada, Phys. Rev. D \textbf{56},
276-290 (1997).

\bibitem{Mizuta}
S.~Mizuta, M.~Yamaguchi, Phys. Lett. B \textbf{298}, 120-126
(1993).

\bibitem{OPAL}
G.~Abbiendi at al. (OPAL Collaboration), CERN-EP/2002-063.

\bibitem{NUSEL}
J. F.~Wilkerson, in "Dark Matter, 2002", Washington University; in
"NIST Physics Colloqium, 2003", Washington University.

\bibitem{aloisio1}
R.~Aloisio, astro-ph/0405110.

\bibitem{gondolo1}
P.~Gondolo, K.~Freese, JHEP 0207, 052 (2002).

\bibitem{gondolo2}
J.~Edsjo, P.~Gondolo, Phys. Rev. D \textbf{56}, 1879-1894 (1997).

\bibitem{pierce}
D.~Pierce, A.~Papadopoulos, Phys. Rev. D \textbf{50}, 565-570
(1994).

\bibitem{Hboson}
M. S.~Chanowitz, Phys. Rev. D \textbf{59}, 073005 (1999).

\bibitem{belyaev}
A.~Belyaev, D.~Garcia, J.~Guasch, JHEP 0206, 059 (2002).

\bibitem{Hagiwara}
Particle Data Group, S.~Eidelman et al., Phys. Lett. B
\textbf{592}, 1 (2004).

\bibitem{Ellis}
J.~Ellis, J.~Hagelin, D.~Nanopoulos, K.~Olive, M.~Srednicki, Nucl.
Phys. B \textbf{238}, 453 (1984).

\bibitem{Jungman}
G.~Jungman, M.~Kamionkowski, K.~Griest, Phys. Rep. \textbf{267},
195 (1996).

\bibitem{gondolo}
P.~Gondolo, astro-ph/0403064.

\bibitem{Primack}
J.~Primack, astro-ph/0112255.

\bibitem{Primack1}
J.~Primack, astro-ph/9707285.

\bibitem{verde}
L.~Verde et al., MNRAS 335, 432 (2002).

\bibitem{spergel}
D. N.~Spergel, et al., Astrophys. J. Suppl. \textbf{148}, 175
(2003).

\bibitem{search1}
R.~Bernabei et al., Riv. Nuov. Cim. 26 n. 1, 1-73 (2003).

\bibitem{smith1}
P. F.~Smith, Phil. Trans. R. Soc. Lond. A \textbf{361}, 2591-2606
(2003).

\bibitem{smith2}
Nigel J. T.~Smith, Nucl. Instr. and Meth. A \textbf{513}, 215-221
(2003).

\bibitem{sumner}
T. J.~Sumner www.livingreviews.org/Articles/Volume5/2002-4sumner

\bibitem{WIMPSearch}
A.~Kurylov, M.~Kamionkowski, Phys. Rev. D \textbf{69}, 063503,
(2004).

\bibitem{ann}
A.~Tasitsiomi, J.~Gaskins, A.~Olinto,  New Astron. Rev.
\textbf{48}, 473-475 (2004).

\bibitem{dremin}
I. M.~Dremin, J. W. Gary, Phys. Rept.  \textbf{349}, 301-393
(2001).

\bibitem{L3}
The L3 Collaboration, Phys. Lett. B \textbf{577}, 109-119 (2003).

\bibitem{hera}
S.~Aid et al., Z. Phys. C \textbf{72}, 573-592 (1996).

\end{thebibliography}
\end{document}